\documentclass[12pt]{article}
\usepackage{latexsym}  % for Symbol fonts
\usepackage{amssymb}
\usepackage{graphicx}
\usepackage{amsmath}
\newcommand{\be}{\begin{equation}}
\newcommand{\ee}{\end{equation}}
\newcommand{\bea}{\begin{eqnarray}}
\newcommand{\eea}{\end{eqnarray}}

\begin{document}
%%%%%%%%%%%%%%%%%%%%%%%%%%%%
\begin{titlepage}
%%%%% PREPRINT NUMBERS %%%%%%
%\begin{flushright}
%\today
%\end{flushright}
%%%%%%%%%%%%%%%%%%%%%%%%%%%%%%
\vspace{3\baselineskip}
%%%%%%%%%%%%%%%%%%% TITLE %%%%%%%%%%%%%%%%%%
\begin{center}
{\Large\bf  Alternative Renormalizable Minimal SO(10) GUT\\ and Seesaw Scale}
\end{center}
%%%%%%%%%%%%%%%% AUTHORS %%%%%%%%%%%%%%%%%%%%%%%
\vspace{1cm}
\begin{center}
%%%%%%%%%%%%%%%%
{\large Takeshi Fukuyama$^{a,}$
\footnote{E-mail:fukuyama@se.ritsumei.ac.jp}}
and
{\large Nobuchika Okada$^{b,}$
\footnote{E-mail:okadan@ua.edu}}
%%%%%%%%%%%%%%%%
\end{center}
%%%%%%%%%%%%%%%%%%%%%%% AFFILIATION %%%%%%%%%%%%
\vspace{0.2cm}
\begin{center}
${}^{a} $ {\small \it  Research Center for Nuclear Physics (RCNP),
Osaka University, \\Ibaraki, Osaka, 567-0047, Japan}\\

${}^{b}$ {\small \it Department of Physics and Astronomy, University of Alabama, \\Tuscaloosa, Alabama 35487, USA} 
\medskip
\vskip 10mm
\end{center}
%\vskip 10mm
%
\begin{abstract}
Alternative renormalizable minimal non-SUSY SO(10) GUT model is proposed. 
Instead of a ${\bf 126}$-dimensional Higgs field, a ${\bf 120}$-dimensional Higgs filed is introduced 
  in addition to a ${\bf 10}$-dimensional Higgs field and plays a crucial role to reproduce 
  the realistic charged fermion mass matrices.  
With contributions of ${\bf 120}$ Higgs field,  the original Witten's scenario of 
  inducing the right-handed Majorana neutrino mass through 2-loop diagrams becomes phenomenologically viable. 
This model inherits the nice features of the conventional renormalizable minimal SO(10) GUT model 
  with ${\bf 10}+{\bf \overline{126}}$ Higgs fields, 
  while supplemented with a low scale seesaw mechanism 
  due to the 2-loop induced right-handed Majorana neutrino mass. 

\end{abstract}
\end{titlepage}
%%%%%%%%%%%%%%%%%%%%%%%%%%%%%%%%%%%%%%%%%%%%%%%%
\section{Introduction}
The great experimental achievements in the recent years, such as the Higgs discovery at the LHC \cite{Aad:2012tfa} 
   and the measurement of the $\theta_{13}$ neutrino mixing angle \cite{An:2012eh} in leptonic mass matrix \cite{Wilking:2013vza}
   can fill the piece of the flavor puzzle and impose a test to the bunch of flavor models.
The supersymmetric grand unified theory (SUSY GUT) can provide the most promising framework 
   to incorporate these vast data systematically and consistently.
Among many candidates, $SO(10)$ \cite{Fritzsch} is the smallest simple gauge group 
  under which the entire Standard Model (SM) matter contents of each generation 
   are unified into a single anomaly-free irreducible representation of ${\bf 16}$. 
The ${\bf 16}$-dimensional spinor representation of the $SO(10)$ gauge group includes 
    the right-handed neutrino but no other exotic matter particles.

A particular attention has been paid to the renormalizable minimal $SO(10)$ model, 
   where two Higgs multiplets  of $\{{\bf 10} \oplus {\bf \overline{126}}\}$-representations 
   are utilized for the Yukawa couplings with the matter fields \cite{Babu:1992ia}. 
The Yukawa couplings to {\bf 10} and $\overline{\bf 126}$ Higgs fields can reproduce realistic charged fermion mass matrices 
   using their phases thoroughly \cite{Matsuda:2000zp, Fukuyama:2002ch, Goh:2003sy}. 
Furthermore, in the minimal SUSY $SO(10)$ model, the renormalizability enables us to construct a superpotential 
   which guides us to the symmetry breaking pattern from the GUT to the SM \cite{clark, FIKMO1} 
   with additional ${\bf 126}$ and ${\bf 210}$-dimensional Higgs fields.

In spite of these nice features, there are deep-rooted opposition to the use of higher dimensional Higgs field 
  like a ${\bf 126}$ representation.
One of the main undesirable feature of this approach is that 
   contributions of such higher dimensional Higgs multiplets to the beta function of the GUT gauge coupling are huge 
   and as a result, the GUT gauge coupling constant very quickly blows up to infinity in its renormalization group (RG) evolution   
   soon after the unification scale ($M_G$). 
For example, in the model with a set of Higgs representations, ${\bf 10 \oplus \overline{126} \oplus 126 \oplus 210}$, 
   the coupling constant diverges at $4.2\times M_G$.
We cannot accept this claim literally since there appear new elements like extra dimensions 
   at around $M_G$ and such naive estimation may not be valid \cite{FO1}. 
However, unless the precedent minimal $SO(10)$ models are fully satisfactory in all directions, 
   it is important to consider possible alternative candidates reserving the fundamental concepts such as minimality and renormalizability.

In non-SUSY case, there is no severe restriction on adopting higher dimensional Higgs fields like ${\bf \overline{126}}$. 
However, in this letter, we try to reproduce the effective ${\bf \overline{126}}$ coupling from loop corrections, 
   which is impossible in the SUSY case because of the non-renormalization theorem \cite{non-ren} for the superpotnetial. 
This was first discussed by Witten \cite{Witten} but found incomplete and unsuccessful in phenomenological point of view. 
In this paper, we reconsider this direction more in detail and shed a new light on a low scale seesaw 
   in the context of the GUT.

This paper is organized as follows: 
In the next section, we review the renormalizable minimal SUSY SO(10) GUT and its nice features. 
Keeping these features as much as possible, we construct an alternative minimal non-SUSY GUT in Section 3. 
The last section is devoted for conclusions.

%%%%%%%%%%%%%%%%%%%%%%%%%%%%%
\section{Renormalizable Minimal SUSY SO(10) GUT}
%%%%%%%%%%%%%%%%%%%%%%%%%%%%%
We briefly review the renormalizable minimal SUSY $SO(10)$ GUT model (minimal $SO(10)$ model) 
   in order to compare it with the alternative non-SUSY one that we will propose in the next section. 
The SM fermion fields are described by a single ${\bf 16}$-dimensional representation in each generation, 
  and the representation of its bilinear is decomposed as 
\be
{\bf 16}\otimes {\bf 16}={\bf 10}\oplus{\bf 120}\oplus{\bf 126}.
\label{1616}
\ee
Hence, Higgs fields are constrained to be in ${\bf 10}$, ${\bf 120}$ and ${\bf \overline{126}}$ representations 
  to make singlets in Yukawa couplings. 
Since a single ${\bf 10}$ Higgs field cannot realize the flavor mixing matrix, 
   at least one more Higgs field needs to be introduced. 
For realistic fermion mass matrices, it is essential to introduce a ${\bf 126}$-dimensional Higgs field.  
Since the ${\bf 126}$-dimensional Higgs field includes $(\overline{{\bf 10}}, {\bf 3}, {\bf 1}) 
 \oplus ({\bf 10}, {\bf 1}, {\bf 3})$ under the subgroup $G_{422}=SU(4)_C\times SU(2)_L\times SU(2)_R$, 
  they play essential roles for generating Majorana masses for left-handed and heavy right-handed neutrinos, respectively.

The Yukawa coupling in the minimal $SO(10)$ GUT is given by 
\begin{eqnarray}
 W_Y = Y_{10}^{ij} {\bf 16}_i {\bf 10}_H {\bf 16}_j 
           +Y_{126}^{ij} {\bf 16}_i {\bf \overline{126}}_H {\bf 16}_j, 
\label{WY}
\end{eqnarray} 
where $i, j$ are the generation indices. 
The ${\bf 10}_H$ and ${\bf \overline{126}}_H$ Higgs fields include a pair of Higgs doublet fields, respectively. 
With vacuum expectation values (VEVs) of these four Higgs doublets, 
   the fermion mass matrices at $M_G$ are described as 
\begin{eqnarray}
&&  M_u = c_{10} M_{10} + c_{126} M_{126}, ~~~~~M_d =     M_{10} +     M_{126},  \nonumber \\ 
&&  M_D = c_{10} M_{10} -3 c_{126} M_{126},~~~M_e =     M_{10} -3     M_{126},   \\
&&  M_R = c_R M_{126}, ~~~~~~~~~~~~~~~~~~~M_L = c_L M_{126}, \nonumber
\label{massmatrix}
\end{eqnarray}
where $c_{10}$, $c_{126}$, $c_R$, and $c_L$ are complex numbers. 
Here, $M_u$, $M_d$, $M_D$, $M_e$, $M_R$, and $M_L$ are mass matrices of 
  up-type quarks, down-type quarks, Dirac neutrinos, charged leptons, right-handed Majorana neutrinos, 
  and left-handed Majorana neutrinos, respectively. 
Through a combination of type I \cite{seesaw} and type II \cite{seesaw2} seesaw mechanism,   
  the light neutrino mass matrix is given by 
\be
M_\nu=M_L-M_D^TM_R^{-1}M_D.
\label{typeII}
\ee
Thus, all matter mass matrices, including the seesaw mechanism ingredients, are given 
  by the two symmetric mass matrices $M_{10}$ and $M_{126}$. 
This is the reason why this model is very predictive for the fermion mass matrices. 
These mass formulas are realized at the GUT scale and the data-fitting is performed 
  after all low energy data of fermion mass matrices are extrapolated to the GUT scale 
  by the RG equations.

%%%%%%%%%%%%%%%%%%%%%%%%%%%%%%%%%%%%%%%
\section{Alternative Renormalizable Minimal SO(10) GUT}
%%%%%%%%%%%%%%%%%%%%%%%%%%%%%%%%%%%%%%%
The requirement that the $SO(10)$ gauge coupling constant remains perturbative up to, say, 
  the Planck scale imposes severe constraint on the set of matter and Higgs representations we can introduce.  
To derive this constraint, we employ the RG evolution of the GUT gauge coupling at the 1-loop level, 
\be 
\frac{1}{\alpha_G(\mu)}=\frac{1}{\alpha_G(M_G)}-\frac{b}{2\pi}\mbox{log}\left(\frac{\mu}{M_G}\right),
\label{RG}
\ee
where $\alpha_G$ is the unified gauge coupling, 
  $b = -b_{gauge} + b_{matter} + b_{Higgs}$ is the beta function coefficient 
  from the contributions due to gauge, matter, and Higgs loops, respectively. 
In the SUSY case, each chiral multiplet contributes $l/2$ to $b$, and each vector (gauge) multiplet contribute $3l/2$, 
   where $l$ is the Dynkin index  of the irreducible representation listed in Table~\ref{table:Dynkin}.
In non-SUSY case, each fermion, scalar and vector fields contribute $l/3$, $l/6$, and $11l/6$, respectively.

%%%%%%%%%%%%%%%%%%
\begin{table}[t]
\centering
\begin{tabular}{|c|c|c|c|c|c|c|c|}
\hline
IRREP & {\bf 10} & {\bf 16} & {\bf 45}  & {\bf 54} & {\bf 120} &{\bf 126}& {\bf 210}\\
\hline
$l/2$ & 1 & 2& 8 & 12 & 28 & 35 & 56 \\
\hline
\end{tabular}
\caption{List of the Dynkin index for $SO(10)$ irreducible representations up to the ${\bf 210}$ dimensional one.}
\label{table:Dynkin}
\end{table}
%%%%%%%%%%%%%%%%%%%%

For SUSY SO(10) models with three families, $b_{gauge}= 24$ and $b_{matter} = 2\times 3$, therefore $b = -18 + b_{Higgs}$. 
For the non-SUSY case, $b_{gauge}= 88/3$ and $b_{matter} = 4/3\times 3$. Then, $b = -18 + b_{Higgs}$ 
  for the SUSY case, while $b = -76/3 + b_{Higgs}$ for the non-SUSY case. 
If we take the constraint and allows the coupling constant to blow up at $\mu=\Lambda$, namely $1/\alpha_G(\Lambda)=0$, we
obtain
\bea
\left(\frac{l}{2}\right)_{Higgs} \leq \left\{
\begin{array}{l}
\displaystyle
18 +\frac{2\pi}{\ln (\frac{\Lambda}{M_G})}\times \frac{1}{\alpha_G(M_G)}~~~~~~~~\mbox{for SUSY case },  \\
\displaystyle
76 +\frac{6\pi}{\ln (\frac{\Lambda}{M_G})}\times \frac{1}{\alpha_G(M_G)}~~~~~~~~\mbox{for non-SUSY case}. 
\end{array}
\right.
\eea
In the RG analysis for the SUSY case, one typically finds $1/\alpha_G(M_G) \simeq 24$ at $M_G=2 \times 10^{16}$ GeV. 
Therefore, $\left(\frac{l}{2}\right)_{Higgs}\leq 49$, if one uses the reduced Planck mass $M_P \simeq 2.4 \times 10^{18}$ GeV 
   for the scale $\Lambda$.  
This constraint excludes the introduction of ${\bf 126}$ and higher dimensional Higgs fields \cite{Chang:2004pb}.   
%If one use the stricter condition $\alpha(M_P ) = 1$, then the constraint becomes $\left(\frac{l}{2}\right)_{Higgs}\leq 48$ 
%  which is about the same as before. 
On the other hand, for non-SUSY SO(10) models, the lower bound on $\left(\frac{l}{2}\right)_{Higgs}$ 
  is much larger than that of the SUSY case, and there is no severe restriction for Higgs multiplets 
  to be introduced. 
Thus, we may consider the non-SUSY version of Eq.~(\ref{WY}) while keeping the GUT gauge coupling  
  in the perturbative regime up to $M_P$. 
However, in the non-SUSY case, there is another way to introduce the right-handed neutrino masses 
  without ${\bf 126}$-dimensional Higgs field at the tree-level, as it was proposed by Witten \cite{Witten}. 
In the following, we focus on this possibility, where the right-handed neutrino masses are loop-induced, and hence 
  the seesaw scale appears at much lower scale than that in usual $SO(10)$ models.

%%%%%%%%%%%%%%%%%%%%%%%%%%%%%%%%%%%%%%%%%%%%%%
\subsection{Loop-induced right-handed neutrino mass with single Yukawa}
%%%%%%%%%%%%%%%%%%%%%%%%%%%%%%%%%%%%%%%%%%%%%%
When an $SO(10)$ model includes $\overline{\bf 126}$ Higgs field, 
   a VEV of the $(\overline{\bf 10},{\bf 1}, {\bf 3})$ component under $G_{422}$ generates the Majorana mass 
   for right-handed neutrinos ($N_R$'s). 
However, even without the $\overline{\bf 126}$ Higgs field, we can generate the Majorana mass 
   when the model includes a ${\bf 16}$-plet Higgs field ($H_{16}$), 
   since a bilinear product of $H_{16}$ can play a role of the $\overline{\bf 126}$ Higgs field. 
An effective operator relevant to this mass generation is given by
\bea 
  {\cal L} \supset \frac{1}{M} {\bf 16}_i  {\bf 16}_j  H_{16}^\dagger  H_{16}^\dagger
\eea  
where $M$ is a mass scale. 
Although we cannot introduce such a higher dimensional term by hand in our renormalizable model, 
  it can be induced through quantum corrections at the 2-loop level, as  has been pointed out by Witten \cite{Witten}. 
This is very interesting since the loop corrections suppress the seesaw scale.

%%%%%%%%%%%%%%%%%%%%%%%%%%%%%%
\begin{figure}[h]
%\centering
%\leavevmode
\begin{center}
\includegraphics[scale=0.2]{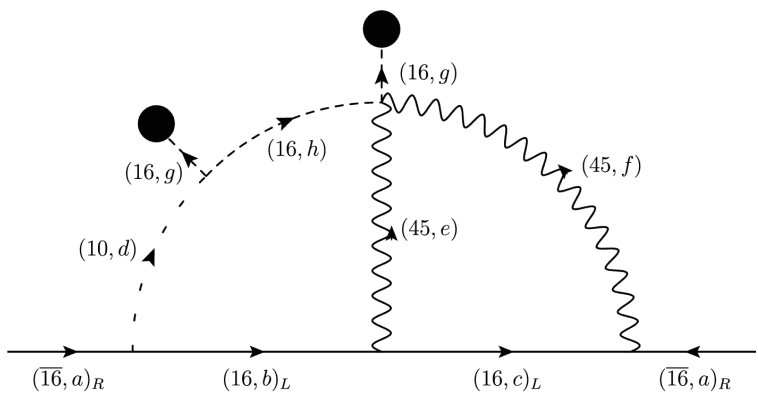}
\end{center}  
%\vspace*{-0.5cm}
\caption{
The construction of the right-handed neutrino mass from
  ${\bf 10} \otimes {\bf 45} \otimes {\bf 45}$ via 2-loop diagram. 
Shown in parenthesis are the $SO(10)$ and the broken subgroup 
 ($SU(5)$ or $G_{422}$) representations, which are summarized in  Table 2. 
Two blobs represent the insertion of $\langle H_{16} \rangle=M_G$.  
The crossed diagram should be added. 
}
\end{figure}
%%%%%%%%%%%%%%%%%%%%%%%%%%%%%%%

%%%%%%%%%%%%%%%%%%%%%%%%%%%%%%%%%
\begin{table}[t]
\centering
\begin{tabular}{|c|c|c|c|c|c|c|c|c|c|}
\hline
   & (a) & (b) & (c)  & (d) & (e) & (f)& (g)& (h) \\
\hline
$SU(5)$ & 1 & $\overline{5}$& 10 & 5 & 10 & 10 & 1& $5$ \\
\hline
$G_{422}$ & 4,1,2 & 4,2,1 & 4,2,1 & 1,2,2& 15,1,1& 6,2,2& 4,1,2& 4,2,1\\
\hline
\end{tabular}
\caption{
Representations of particles in the 2-loop diagram under the $SO(10)$ subgroups. 
}
\end{table}
%%%%%%%%%%%%%%%%%%%%%%%%%%%%%%%%%%%

In the simplest model discussed in Ref.~\cite{Witten}, the matter fermions couple directly  only to ${\bf 10}$ Higgs like in Eq.~(\ref{WY}) and the $SO(10)$ gauge field of the ${\bf 45}$ representation. 
The basic idea is that ${\bf 126}$ representation is a 5th rank tensor 
  which can be constructed by ${\bf 10} \otimes {\bf 45} \otimes {\bf 45}$ 
  with a vector ${\bf 10}$ and a 2nd rank tensor ${\bf 45}$. 
In fact, the $N_R$ mass is generated by quantum corrections at the 2-loop level 
  as shown in Fig.~1 \cite{Witten}, when $H_{16}$ develops its VEV to break the $SO(10)$ group 
  to its subgroup $SU(5)$. 
Here, note that a triple scalar coupling among the Higgs fields also plays a crucial role: 
\bea 
  {\cal L} \supset \lambda_{10} M_G H_{10} H_{16} H_{16},   
\label{3-scalar}  
\eea
 where we have parametrize the triple scalar coupling with $M_G$ and 
 a dimensionless coupling constant $\lambda_{10}$. 
The resultant $N_R$ mass is estimated as \cite{Witten}  
\bea
M_R = \left( \frac{m_q}{M_W} \right) \epsilon_{10} \left( \frac{\alpha_G}{\pi} \right)^2 M_G.
\label{MR}
\eea
Here, we have used a relation,  $Y_{10} \sim m_q/M_W$
   between the Yukawa coupling of $H_{10}$ and an up-type quark mass $m_q$, 
   and $\epsilon_{10}$ represents a mixing angle between $H_{10}$ and $H_{16}$ 
   induced by their coupling in Eq.~(\ref{3-scalar}) with a VEV of $H_{16}$.    
Note that the $M_R$ scale is much lower than the usual seesaw scale of the model $\sim M_G$.    
In the present $SO(10)$ model with only one Yukawa coupling $Y_{10}$, 
  the Dirac neutrino mass matrix is the same as the up-type quark mass matrix, 
  and therefore the light neutrino mass due to the type I seesaw mechanism 
  (see the 2nd term in Eq.~(\ref{typeII}) in the right-handed side) is estimated as  
\bea
m_{\nu_L} = m_q  \left[ \epsilon \left( \frac{\alpha_G}{\pi} \right)^2 \right]^{-1} \frac{M_W}{M_G}.
\label{mnu}
\eea
As in Ref.~\cite{Witten}, we estimate $m_{\nu_L}=10^{-7} \, m_q$ 
  by using $(\alpha_G/\pi)^2 =10^{-5}$, $M_G=10^{15}$ GeV and $\epsilon=0.1$. 
Clearly, the light neutrino mass spectrum predicted by this formula is unrealistic. 
For example, the heaviest light neutrino mass is $10^{-7}\, m_t \sim 20$ keV, 
   where $m_t=173$ GeV is the top quark mass.

Let us here recall the situation in the minimal SO(10) GUT for the fitting of the fermion mass matrices. 
In Eq.~(\ref{massmatrix}), the fermion mass matrices 
  are expressed by two mass matrices $M_{10}$ and $M_{126}$. 
A good relation of $m_\tau \approx m_b$ at the GUT scale 
  is satisfied when $M_{10}$ dominates over $M_{126}$ in the 3rd generation, 
  while such a relation is not suitable for the mass relations between charged leptons and 
  down-type quarks in the 1st and 2nd generations. 
For the mass fitting for the leptons and quarks in 1st and 2nd generations, 
  a delicate tuning between $M_{10}$ and $M_{126}$  is crucial in the minimal $SO(10)$ GUT. 
We may expect a similar parameter tuning  to reduce the too large mass eigenvalues 
  for the light neutrinos. 
For this purpose, we consider type II seesaw extension of the model. 
It is useful to express the representations of the particles in the 2-loop diagram 
  in terms of the subgroup $G_{422}$ as listed by the last row in Table 2. 
Note that if we replace the role of $SU(2)_R$ into $SU(2)_L$, 
  the 2-loop diagram in Fig.~1 is a diagram to generate $M_L$ in Eq.~(\ref{typeII}) 
  with an $SU(2)_L$ doublet VEV ($v_{EW}$) in $H_{16}$: 
\bea
M_L=\left( \frac{m_q}{M_W} \right) \epsilon \left( \frac{\alpha_G}{\pi} \right)^2 \frac{v_{EW}^{~~~2}}{M_G}.
\label{ML}
\eea
Then, we may expect a cancellation between the two terms in the right-hand side of Eq.~(\ref{typeII}), 
  to reduce the mass scale of $M_\nu$. 
Unfortunately, this cancellation cannot work, because $v_{EW}$ is at most the the electroweak scale 
  and the scale of $M_L$ is too small.

The too large mass scale of $m_{\nu_{Li}}$ originates from the quark mass ($m_q$) insertions 
   in Eqs.~(\ref{MR}) and (\ref{mnu}). 
This is natural in the $SO(10)$ model since all fermions of the SM in each generation 
  are unified into a single ${\bf 16}$ representation. 
We may consider a partial GUT model, instead of the $SO(10)$ group, 
  where  the Yukawa couplings of quarks and leptons are not unified and thus independent. 
A simple model based on a subgroup of $SO(10)$ is a Left-Right symmetric model 
  with the gauge group $G_{3221}=SU(3)_C \otimes SU(2)_L\otimes SU(2)_R\otimes U(1)_{B-L}$.  
If we can replace $m_q$ in Eqs.~(\ref{MR}) and (\ref{mnu}) into a charged lepton mass ($m_\ell$), 
  we obtain 
\bea
M_R = \left( \frac{m_\ell}{M_W} \right) \epsilon \left( \frac{\alpha_G}{\pi} \right)^2 M_G.
\label{MR2}
\eea
and
\bea
m_{\nu_L} = m_\ell  \left[ \epsilon \left( \frac{\alpha_G}{\pi} \right)^2 \right]^{-1} \frac{M_W}{M_G}.
\label{mnu2}
\eea
Now, in estimating the heaviest light neutrino mass, 
   we use $m_\tau=1.78$ GeV instead of $m_t=173$ GeV and obtain $m_{\nu_\tau} \sim 100$ eV. 
If we can choose $M_G =M_P=2.4 \times 10^{18}$ GeV, 
  we obtain $m_{\nu_\tau} = {\cal O}$(0.1 eV), and thus the situation is  much improved. 
Since the light neutrino mass spectrum is determined by the charged lepton masses, 
  we find  
\bea
 \frac{\Delta m_{23}^2}{\Delta m_{12}^2} = \left( \frac{m_\tau}{m_\mu}\right)^2 \simeq 283. 
\eea
This should be compared with the experimental value, $\Delta m_{23}^2/\Delta m_{12}^2 \simeq 32$ 
  for $\Delta m_{12}^2=7.6 \times 10^{-5}$ eV$^2$ and $\Delta m_{23}^2=2.4 \times 10^{-3}$ eV$^2$ \cite{PDG}. 
The predicted ratio is not perfect but not very bad. 
However, the 2-loop diagram in Fig.~1 cannot be constructed in the Left-Right symmetric model. 
We can see this fact from Table 2. 
While the 2-loop diagram includes the gauge boson in $(6,2,2)$ representation of the subgroup $G_{422}$ 
 (see the column (f)), it is not included in the Left-Right symmetric model.

%%%%%%%%%%%%%%%%%%%%%%%%%%%%%%%%%
\subsection{Introduction to {\bf 120}-dimensional Higgs field}
%%%%%%%%%%%%%%%%%%%%%%%%%%%%%%%%%
In the previous subsection, we have shown that the light neutrino mass matrix 
  from type I seesaw with the 2-loop induced $M_R$ is unrealistic. 
This is because of the quark mass $m_q$ insertions in Eqs.~(\ref{MR}) and (\ref{mnu}), 
  which originate from the single Yukawa coupling $Y_{10}$. 
A simple way to ameliorate the problem is to add one more Yukawa coupling.  
Note that this is, in fact, necessary not only for a realistic neutrino mass matrix, 
  but also realistic charged fermion mass matrices. 
Recall from Eq.~(\ref{massmatrix}) that the mass matrix relation with only $M_{10}$ 
  is clearly unrealistic since $M_u \propto M_d = M_e$. 
The nice feature of the minimal $SO(10)$ model is that this bad mass matrix relation 
  is corrected by introducing the $\overline{\bf 126}$ Higgs field.  
For a minimal setup alternative to the minimal $SO(10)$ model, 
  we introduce ${\bf 120}$-dimensional Higgs field. 
Although the ${\bf 120}$ Higgs field includes two $SU(2)_L$ Higgs doublets like the $\overline{\bf 126}$ Higgs field, 
  it does not involve $(\overline{{\bf 10}}, {\bf 3}, {\bf 1}) \oplus ({\bf 10}, {\bf 1}, {\bf 3})$ under the subgroup $G_{422}$.  
Hence, the Majorana neutrino mass matrices are not generated at the tree-level.

With the ${\bf 120}$ Higgs field, the Yukawa interactions are given by    
\bea
 {\cal L}_Y = Y_{10}^{ij} {\bf 16}_i {\bf 10}_H {\bf 16}_j 
           +Y_{120}^{ij} {\bf 16}_i {\bf 120}_H {\bf 16}_j.  
\label{Y_120}
\eea
With VEVs of four Higgs doublets (two in ${\bf 10}_H$ and the other two in ${\bf 120}_H$), 
  the fermion mass matrices at $M_G$ are described as 
\begin{eqnarray}
&&  M_u = c_{10} M_{10} + c_{120} M_{120}, ~~~~~M_d =     M_{10} +     M_{120},  \nonumber \\ 
&&  M_D = c_{10} M_{10} -3 c_{120} M_{120},~~~M_e =     M_{10} -3     M_{120},   
\label{massmatrix_120}
\end{eqnarray}
where $c_{10}$ are $c_{120}$ are complex numbers. 
This mass matrix relation is similar to that in the minimal $SO(10)$ model. 
It has been shown in Ref.~\cite{Matsuda:2000zp} (see also Ref.~\cite{Chang:2004pb}) 
  that the system with $\left\{{\bf 10} \oplus {\bf 120} \right\}$ Higgs fields can reproduce the realistic charged 
  fermion mass matrices.

One may thank that the ${\bf 120}$ Higgs field also contributes to $M_R$ through quantum corrections at the 2-loop level. 
Similarly to the ${\bf 10}$ Higgs case, we may introduce a triple scalar coupling 
\bea 
  {\cal L} \supset \lambda_{120} M_G H_{120} H_{16} H_{16}.   
\label{3-scalar-120}  
\eea
However, since $H_{120}$ is an antisymmetric tensor, this term is vanishing. 
Thus, there is no new contribution to $M_R$ from $H_{120}$. 
%Then, the 2-loop diagram is the same as Fig.~1, but $(10, d)$ is replaced to $(120, d)$, 
%  while the representations in Table 2 are the same.  
%The resultant $N_R$ mass is now estimated as 
%\bea
%M_R = \left( Y_{10} \epsilon_{10} + Y_{120} \epsilon_{120} \right)  \left( \frac{\alpha_G}{\pi} \right)^2 M_G, 
%\label{MR_120}
%\eea
%where $\epsilon_{120}$ represents a mixing angle between $H_{120}$ and $H_{16}$ 
%   induced by their coupling in Eq.~(\ref{3-scalar-120}) with a VEV of $H_{16}$.    
As in Eq.~(\ref{ML}), a new contribution to $M_L$ with the ${\bf 120}$ Higgs is generated, 
  but we neglect it since it is extremely small. 
It may be useful to express Eq.~(\ref{MR}) in terms of $M_{10}$ as
\bea
 M_R = c_R^{10} M_{10} ,
\label{MR_120-2}  
\eea 
 where the coefficient is given by
\bea
 c_R^{10} \sim  \epsilon_{10}  \left( \frac{\alpha_G}{\pi} \right)^2 \frac{M_G}{M_W} . 
 %, \; \; \; 
% c_R^{120} \sim  \epsilon_{120}  \left( \frac{\alpha_G}{\pi} \right)^2 \frac{M_G}{M_W} . 
\eea 
The crucial difference from the single Yukawa case in the previous subsection is that 
  $M_D$ is a linear combination of $M_{10}$ and $M_{120}$. 
If a cancellation between $c_{10} M_{10}$ and $3 c_{120} M_{120}$ occurs,  
   the scale of $M_D$ becomes lower than the single Yukawa case.  
As a result, there is a possibility to circumvent the pathology in the single Yukawa case and 
   the resultant mass scale of light neutrino mass matrix can be of order $0.1$ eV.

%%%%%%%%%%%%%%%%%
\section{Conclusions}
%%%%%%%%%%%%%%%%%
We have considered a non-SUSY renormalizable $SO(10)$ GUT without $\overline{\bf 126}$ Higgs field, 
   where the right-handed Majorana neutrino mass is induced at the 2-loop level. 
The case with a single Yukawa coupling $Y_{10}$, which is originally discussed by Witten in Ref.~\cite{Witten}, 
   is in any case unrealistic in terms of not only the light neutrino mass matrix but also the charged fermion mass matrices. 
In order to solve this problem, we have extended the Higgs sector by adding one ${\bf 120}$ Higgs field. 
Such an extension is crucial to reproduce the realistic charged fermion mass matrices. 
At the same time, the ${\bf 120}$ Higgs field 
%gives rise to additional loop-contributions to the right-handed Majorana neutrino mass matrix and 
   opens up a possibility to achieve the realistic light neutrino mass matrix after type I seesaw 
   with a possible cancellation between $c_{10} M_{10}$ and $3 c_{120} M_{120}$. 
Our model with the ${\bf 10}+{\bf 120}$ Higgs fields inherits the advantageous points of the conventional 
   minimal $SO(10)$ model with ${\bf 10}+{\bf \overline{126}}$ Higgs fields, while supplemented 
   with a low scale seesaw due to the 2-loop induced right-handed Majorana neutrino mass matrix. 
Since our model involves only 2 Higgs fields for the Yukawa couplings just like the minimal $SO(10)$ model, 
   our model is also the minimal model alternative to the conventional one. 
In order to show a phenomenological viability of the model,  
  the detailed fitting for the fermion mass matrices with up-to-date experimental data is necessary, 
  which will be addressed in the future work \cite{FOT}.

%%%%%%%%%%%%%%%%%%%
\section*{Acknowledgments}
%%%%%%%%%%%%%%%%%%%

This work is supported in part by 
  Grant-in-Aid for Science Research from the Ministry of Education, Science and Culture No.~17H01133 (T.F.) 
  and the U.S. Department of Energy No.~DE-SE0012447 (N.O.).

\end{document}